\documentclass[preprint,prb,superscriptaddress,showpacs]{revtex4}

\usepackage{amssymb, amsmath}

\usepackage{graphicx}
\usepackage{color}

\begin{document}

\title{Magnetic phase transitions in single crystals of the chiral helimagnet Cr$_{1/3}$NbS$_{2}$}

\author{N. J. Ghimire}


\affiliation{Department of Physics and Astronomy, The University of Tennessee, Knoxville, Tennessee 37996, USA}

\affiliation{Materials Science and Technology Division, Oak Ridge National Laboratory, Oak Ridge, Tennessee 37831, USA}

\author{M. A. McGuire}


\affiliation{Materials Science and Technology Division, Oak Ridge National Laboratory, Oak Ridge, Tennessee 37831, USA}

\author{D. S. Parker}


\affiliation{Materials Science and Technology Division, Oak Ridge National Laboratory, Oak Ridge, Tennessee 37831, USA}

\author{B. Sipos}


\affiliation{Materials Science and Technology Division, Oak Ridge National Laboratory, Oak Ridge, Tennessee 37831, USA}

\author{S. Tang}


\affiliation{Materials Science and Technology Division, Oak Ridge National Laboratory, Oak Ridge, Tennessee 37831, USA}

\author{J.-Q. Yan}


\affiliation{Materials Science and Technology Division, Oak Ridge National Laboratory, Oak Ridge, Tennessee 37831, USA}
\affiliation{Department of Materials Science and Engineering, The University of Tennessee, Knoxville, Tennessee, 37996, USA}

\author{B. C. Sales}


\affiliation{Materials Science and Technology Division, Oak Ridge National Laboratory, Oak Ridge, Tennessee 37831, USA}
\author{D. Mandrus}


\affiliation{Department of Physics and Astronomy, The University of Tennessee, Knoxville, Tennessee 37996, USA}
\affiliation{Materials Science and Technology Division, Oak Ridge National Laboratory, Oak Ridge, Tennessee 37831, USA}
\affiliation{Department of Materials Science and Engineering, The University of Tennessee, Knoxville, Tennessee, 37996, USA}

\date{\today}

\begin{abstract}
The chiral helimagnet Cr$_{1/3}$NbS$_{2}$ has been investigated by magnetic, transport and thermal properties measurements on single crystals and by first principles electronic structure calculations. From the measured field and temperature dependence of the magnetization for fields applied perpendicular to the \emph{c} axis, the magnetic phase diagram has been constructed in the vicinity of the phase transitions. A transition from a paramagnetic to a magnetically ordered phase occurs near 120 K. With increasing magnetic field and at temperatures below 120 K, this material undergoes transitions from a helimagnetic to a soliton-lattice phase near 900 Oe, and then to a ferromagnetic phase near 1300 Oe. The transitions are found to strongly affect the electrical transport. The resistivity decreases sharply upon cooling near 120 K, and the spin reorientation from the helimagnetic ground state to the commensurate ferromagnetic state is evident in the magnetoresistance. At high fields a large magnetoresistance (55 \% at 140 kOe) is observed near the magnetic transition temperature. Heat capacity and electronic structure calculations show the density of states at the Fermi level is low in the magnetically ordered state. Effects of spin fluctuations are likely important in understanding the behavior of Cr$_{1/3}$NbS$_{2}$ near and above the magnetic ordering transitions.
\end{abstract}
\maketitle

\section{Introduction}
Complex magnetic textures have recently been observed in some magnetic materials crystallizing in the cubic space group P2${_1}$3. \cite{Muhlbauer2009a, Yu2011a, Yu2010, Munzer2010, Seki2012} MnSi, Fe$_{0.5}$Co$_{0.5}$Si, FeGe, and Cu$_{2}$OSeO$_{3}$ are those materials where the ground state magnetic structure is found to have a modulated helimagnetic arrangement of spins incommensurate with the underlying crystal lattice. In addition to this, another remarkable similarity in these materials is that in a certain range of magnetic fields and at temperatures just below the transition temperature, a distinct skyrmion lattice phase exists. A skyrmion is a magnetic vortex, a structure reminiscent of the Abrikosov vortices in type II superconductors.\cite{Bogadonov1989c,Bogdanov2002,Rossler2006,Rossler2011} It is to be noted that these similarities have been observed irrespective of the nature of the electrical conductivity of the materials. MnSi and FeGe are metals, Fe$_{0.5}$Co$_{0.5}$Si is a small band gap semiconductor  and Cu$_{2}$OSeO$_{3}$ is an insulator. The first three have ferromagnetic ordering and the latter one has ferrimagnetic ordering in a certain plane that modulates along a particular crystallographic direction to form a helical ground state. The helix eventually gets destabilized to long ranged ferromagnetic and ferrimagnetic ordering respectively with the application of magnetic field. A common property all the materials share is the symmetry of the crystal structure. The remarkable feature of space group P2${_1}$3 is that it lacks an inversion center. The ground state chiral helimagnetic ordering in the non-centrosymmetric space group has been well understood to be due to the competition between the symmetric exchange interaction that favors a parallel arrangement of spins and the antisymmetric Dzyaloshinsky-Moriya(DM) interaction that favors a perpendicular arrangement of spins, which results in a modulated helimagnetic spin structure. \cite{Bak1980, Nakanishi1980b, Kataoka1981, Kataoka1984} The origin of the skyrmion phase, however, is still unclear on a general level. Theoretically, it has been predicted to occur in a wide range of non-centrosymmetric magnets. \cite{Bogadonov1989c, Bogdanov1989, Bogdanov1994} However, experimentally they have been observed in materials belonging to only one space group. At the same time, recent studies have shown that the non-centrosymmetric magnets provide a platform for the observation of a magnetic blue phase, a novel magnetic state, formed by the twisting of the chiral helices, reminiscent of the blue phase observed in nematic liquid crystals. \cite{Hamann2011,Fischer2008,Wright1989}

Cr$_{1/3}$NbS$_{2}$ is a material that has all the common properties of the materials mentioned above except that it bears a different crystallographic symmetry. It crystallizes in a noncentrosymmetric hexagonal space group P6$_{3}$22 belonging to the point group D$_{6}^{6}$, \cite{VanLaar1971a} one of the symmetry groups in which ferromagnets have been theoretically predicted to have a skyrmion lattice. \cite{Bogdanov1994} It orders magnetically into a helimagnetic ground state with a period of about 480 \AA\ below about 120 - 127 K and is known to have metallic behavior below room temperature. \cite{Hulliger1970, Miyadai1983, Parkin1980b, Parkin1980c} The spins are arranged ferromagnetically in the \emph{ab}-plane and the helix is along the \emph{c} axis. Effect of magnetic field in the \emph{ab}-plane has been found to be dramatic with a metamagnetic transition observed at an applied field near 1200-1500 Oe. \cite{Miyadai1983} A recent study conducted on a thin crystal with small angle electron diffraction and Lorentz force microscopy has shown that a magnetic field applied perpendicular to the direction of the helix destabilizes the helical structure gradually into a soliton lattice, a nonlinear periodic magnetic state, with an eventual incommensurate-to-commensurate transition into a ferromagnetic state at the critical field of 2300 Oe. \cite{Togawa2012} Manipulation of the spin spiral with magnetic field has generated interest in this material for spintronics applications.\cite{Kishine2009,Kishine2011,Kiselev2011,Bostrem2008}

The first helimagnetic ground state in this material was experimentally observed by Miyadai \emph{et al.} \cite{Miyadai1983} by means of small angle neutron scattering conducted on a powder sample,  which was in accordance with the prediction of the helimagnetic structure by Moriya and Miyadai \cite{Moriya1982} based on the magnetic measurements and the subsequent theoretical interpretation in terms of the DM interaction in a non-centrosymmetric magnet. In earlier studies Parkin \emph{et al.} \cite{Parkin1980c, Parkin1980b} found it to order ferromagnetically below 120 K with the basal plane being the easy axis.

In this paper we present experimental results obtained on high quality single crystals of Cr$_{1/3}$NbS$_{2}$ together with  results from electronic structure calculations. Temperature dependence of DC susceptibility and the magnetization curves are measured both parallel and perpendicular to the \emph{c} axis. Likewise, the temperature dependence of electrical resistivity, specific heat capacity, thermal conductivity and Seebeck coefficient in zero magnetic field and their behavior in magnetic fields applied perpendicular to the \emph{c} axis are investigated. Results of band structure and density of states (DOS) calculations are presented both in the non-magnetic and magnetic states.
\section{Crystal Chemistry}
\begin{figure}[h]
\begin{center}
\includegraphics[scale=1.5]{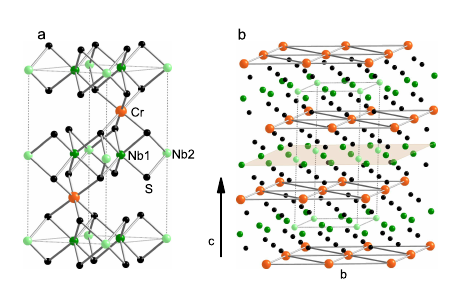}
\caption{\label{1a}(Color online) a) Hexagonal crystal structure of Cr$_{1/3}$NbS$_{2}$. The intercalated chromium atoms occupy the octahedral interstitial holes between the trigonal prismatic layers of 2H-NbS2. b) Structure of Cr$_{1/3}$NbS$_{2}$ emphasizing the layers of Cr atoms in the \emph{ab}-plane. The Cr-Cr distance is shortest in the plane. The dotted lines show the unit cell. The biggest (orange)  balls represent Cr atoms, the medium (green) balls  represent Nb atoms and the smallest (black) balls represent S atoms. Nb atoms are in two inequivalent sites labeled as Nb1 and Nb2.}
\end{center}
\end{figure}

Cr$_{1/3}$NbS$_{2}$ crystallizes in the Nb$_{3}$CoS$_{6}$, hp20 structure type. \cite{Hulliger1970} It forms a hexagonal layered structure as depicted in Figure \ref{1a}(a) containing of 20 atoms per unit cell. Twelve sulfur atoms occupy the general site. There are six niobium atoms in two inequivalent positions - 4f and 2a.  The Cr atoms are intercalated in the octahedral holes (2c sites) between the trigonal prismatic layers of 2H-NbS$_{2}$. The unit cell parameters obtained from the x-ray data collected from the powdered crystals are \emph{a} = 5.741 \AA\ and \emph{c} = 12.101 \AA\ which are in good agreement with the reported values. \cite{Hulliger1970, Miyadai1983} The Cr-Cr distance is closest within the layer in the \emph{ab}-plane (5.741 \AA). Along the \emph{c} axis, the nearest Cr-Cr distance is 6.847 \AA. Cr$_{1/3}$NbS$_{2}$ belongs to a large family of layered materials that are formed by intercalation of transition metal elements within the layers of the first row transition metal dichalcogenides. The structure depends on the amount of intercalation in the layer. In the absence of intercalated atoms the host transition metal dichalcogenide layers are coupled through weak van der Waals bond. The intercalation of a transition metal strengthens the bonding with the possibility of charge transfer from the intercalated atoms to the transition metal atoms in the layer which brings about strong changes in the electronic structure and corresponding changes in the electrical transport and magnetic properties.\cite{Beal1979}

\section{Experimental details}
Crystals of Cr$_{1/3}$NbS$_{2}$ were grown by chemical vapor transport using iodine as the transport agent \cite{Miyadai1983}. Single crystals of various sizes with dimensions as large as 8 mm $\times$ 7 mm $\times$ 1 mm were obtained. X-ray diffraction from powdered crystals confirmed the P6$_{3}$22 structure and revealed no impurity phases. Energy dispersive x-ray spectrometer (EDS) results along with the magnetic transition temperature clearly distinguish these crystals as having 1/3 intercalation of Cr atoms. \cite{Parkin1980a, Parkin1980b} No iodine was detected in the crystals by the EDS measurements. Magnetic properties were measured using a Quantum Design magnetic property measurement system (MPMS). Resistivity, magnetoresistance, thermal conductivity and Seebeck coefficient were all measured in a Quantum Design physical property measurement system (PPMS).  A four wire configuration with platinum wires and Epotek H20E silver epoxy were used for the the resistivity and magnetoresistance measurements.

\section{Results and Discussion}
\subsection{Magnetic properties}
\begin{figure}[h]
\begin{center}
\includegraphics[scale=1.2]{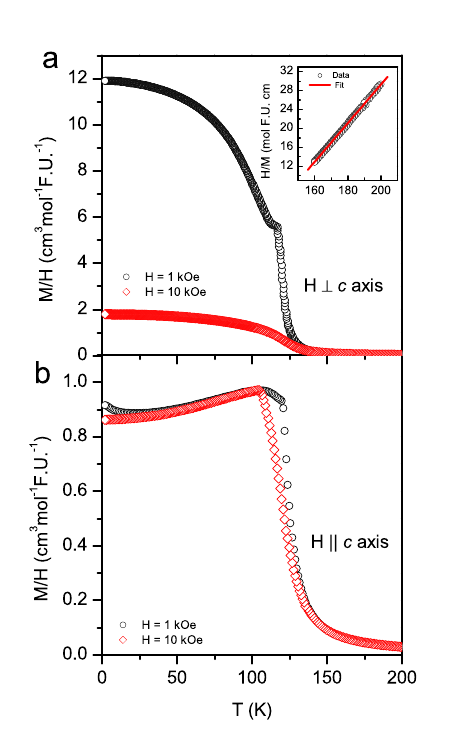}
\caption{\label{1}(Color online) $M/H$  as a function of temperature at the fields indicated with the magnetic field applied a) perpendicular to the \emph{c} axis, and b) parallel to the \emph{c} axis. Inset in (a) shows the fit to the Curie-Weiss law for the data taken at 1 kOe above 160 K.}
\end{center}
\end{figure}

 Figure \ref{1} shows \emph{M/H} as a function of temperature with the applied magnetic field perpendicular [Fig. \ref{1}(a)] and parallel [Fig. \ref{1}(b)] to the \emph{c} axis.  With the magnetic field applied perpendicular to the \emph{c} axis, at the lower field (1 kOe), a small kink is observed at about 117 K. This kink is not observed at 10 kOe. No such kink is observed in \emph{M/H} vs. \emph{T} when the magnetic field is applied parallel to the \emph{c} axis. These results are consistent with previous measurements by Miyadai \emph{et al}. \cite{Miyadai1983} As the material is known to have helimagnetic ordering in lower fields with the helix directed along the \emph{c} axis, \cite{Togawa2012} the kink may represent the onset of the helical state. Such a kink has also been observed in other known helimagnets MnSi \cite{Chattopadhyay2009} FeGe \cite{Ludgren1970} and Fe$_{1-x}$Co$_{x}$Si \cite{Onose2005}. Cr$_{1/3}$NbS$_{2}$ follows Curie-Weiss behavior at higher temperatures. The inset in Fig. \ref{1}(a) shows the Curie-Weiss fit of $\chi^{-1} = \frac{C}{T-\theta_{cw}}$  to the high temperature part of data measured with the applied field of 1 kOe. The parameters obtained are the Curie constant \emph{C} = 2.4 K cm$^{3}$ mol $^{-1}$ F.U.$^{-1}$ and the Curie-Weiss temperature $\theta_{CW}$ = 127 K. The effective moment per mole of chromium atoms, $p_{eff}$, calculated from the Curie constant is 4.4 $\mu_{B}$, which is consistent with the values reported by Parkin \emph{et al.} \cite{Parkin1980a} and is close to the spin only value of 3.87 $\mu_{B}$ for Cr$^{3+}$.
\begin{figure}[h]
\begin{center}
\includegraphics[scale=1.2]{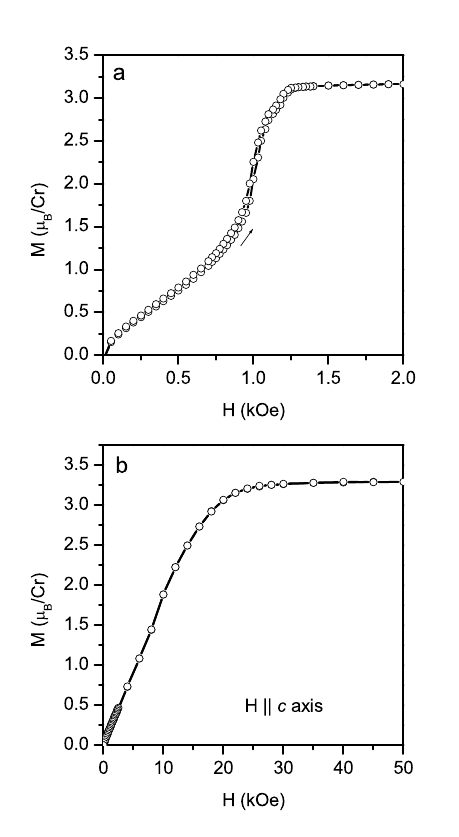}
\caption{\label{2} $M$ vs. $H$  measured at a temperature of 2 K with the magnetic field applied a) perpendicular, and b) parallel to the \emph{c} axis. Note the different scales on the x-axis.}
\end{center}
\end{figure}

The anisotropic nature of the low temperature magnetism in Cr$_{1/3}$NbS$_{2}$ is evident in Fig. \ref{2}. Applying a field along the \emph{c} axis [Fig. \ref{2}(b)] simply rotates the ordered moments out of the \emph{ab}-plane. This results in a nearly linear increase in \emph{M} with \emph{H} up to \emph{H} = 20 kOe. When the field is applied in the \emph{ab}-plane [Fig. \ref{2}(a)], more complex behavior is observed, which will be addressed further in the following discussion. The moment saturates to the same value of about 3.2 $\mu_{B}$/Cr in both orientations, but this occurs at a much lower field when \emph{H} is applied in the \emph{ab}-plane, indicating a strong preference for the moments to remain perpendicular to the \emph{c} axis.
\begin{figure}[h]
\begin{center}
\includegraphics[scale=1.0]{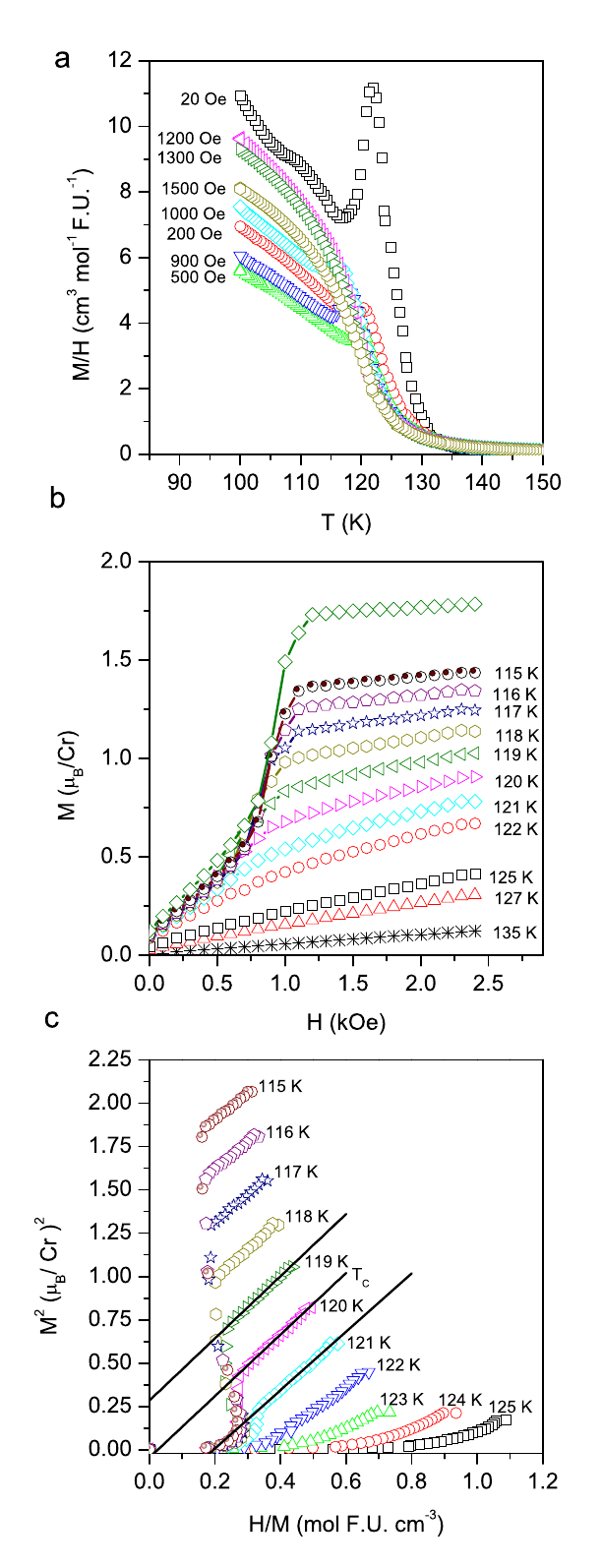}
\caption{\label{2b} (Color online) Magnetic properties of Cr$_{1/3}$NbS$_{2}$ near the transition temperature in low magnetic field applied perpendicular to the \emph{c} axis. a) $M/H$ as a function of temperature at indicated fields. b) $M$ vs. $H$ at selected temperatures, and c) Arrott plots. The solid lines are the extrapolation of the $M^{2}$ vs. $M/H$ data at higher fields.}
\end{center}
\end{figure}

Figure \ref{2b} shows various magnetic properties near the magnetic ordering temperature with the magnetic field applied perpendicular to the \emph{c} axis i.e. perpendicular to the helical axis. \emph{M/H} as a function of temperature at the indicated magnetic fields are depicted in Fig. \ref{2b}(a). It is observed that the kink due to the onset of helimagnetic ordering vanishes above 1100 Oe. At 119 K, \emph{M} vs. \emph{H} increases linearly at low field. At around 500 Oe, it shows a change in slope. Another slope change is observed around 1000 Oe. The region between 500 - 1000 Oe shows a steeper slope. With the decrease in temperature, the lower end of the steeper slope moves towards higher field and a sharp change is observed around 1000 Oe. The moments saturate above about 1300 Oe.  These observations in the magnetic measurements are consistent with the helimagnetic ordering and soliton lattice phase formation in the material as observed by Togawa \emph{et al}. \cite{Togawa2012} In zero applied field, below the transition temperature the moments in a plane perpendicular to \emph{c} axis align parallel to each other while the moments in the plane make small angle to the parallel moments in the adjacent planes. The kink thus represents the effect of the small antisymmetric coupling between the planes giving rise to a long period helimagnetic ordering. As an external magnetic field is applied perpendicular to the \emph{c} axis, at smaller fields (up to 500 Oe) the field is not strong enough to overcome the antisymmetric coupling between the planes. It can, however, turn the moments in the planes gradually in the direction of the field. Thus, \emph{M} increases slowly with \emph{H} at low fields. As the field gets stronger, it starts breaking the antisymmetric coupling between the planes and the moments in a certain plane start to rotate in the direction of the applied field. This results in a rapid increase of \emph{M}. This region of the phase diagram is where the period of the helix is found to increase with the  applied magnetic field forming the soliton phase. \cite{Togawa2012} A large enough field (1300 Oe) aligns all the moments parallel to the applied field thus causing the transition to the ferromagnetic state and saturation of the moment.

\begin{figure}[h]
\begin{center}
\includegraphics[scale=1.7]{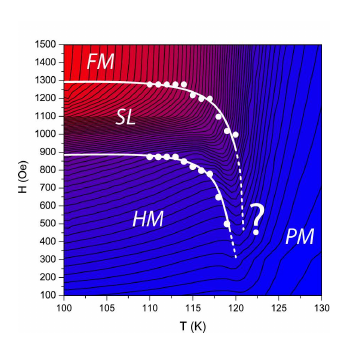}
\caption{\label{2c} (Color online) Contours of \emph{M (H,T)} determined from \emph{M} vs.\emph{T} measurements at fields from 100 Oe to 1500 Oe applied in the \emph{ab}-plane. The white data points were determined from \emph{M} vs. \emph{H} measurements at temperatures from 110 to 120 K. The question mark ``?" represents the region of transition form paramagnetic (PM) to helimagnetic (HM) phase. The details of this phase evolution have not yet been determined.}
\end{center}
\end{figure}

Figure \ref{2c} shows a contour plot of \emph{M} vs. \emph{H} and \emph{T} near the phase transition region. This provides an useful visualization of the magnetic phase diagram for Cr$_{1/3}$NbS$_{2}$. The white lines are meant to guide the eye, and follow the temperature and field dependence of the transition temperatures (white circles) determined from examination of individual \emph{M(H)} curves at fixed \emph{T} [see, for example, the 110 K curve in Fig. \ref{2b}(b)]. At high temperature, Cr$_{1/3}$NbS$_{2}$ is paramagnetic (PM). The lower set of points separates the helimagnetic (HM) phase from the soliton lattice (SL) phase, and is determined by the onset of the rapid increase in \emph{M} with \emph{H}. The upper set of points separates the SL phase from the ferromagnetic (FM) phase, and is determined by the saturation field at a particular temperature. The details of how the complex HM and SL phases evolve upon cooling from the PM phase have not been determined. Here we observed the transition from the soliton phase to the ferromagnetic state at field of about 1300 Oe. In the thin sample measured by Lorentz microscopy, Togawa \emph{et al} \cite{Togawa2012} found it to occur at 2300 Oe. In our crystals, the manifestation of the metamagnetic transition is apparent even in the thermal and transport properties discussed below.

Figure \ref{2b}(c) shows variation of $M^{2}(H,T)$ as a function of $H/M(H,T)$ at different temperatures in the region of the magnetic transition. These Arrott plots in the ferromagnetic state (at higher fields) are linear, indicating itinerant ferromagnetism. \cite{Arrott1967,Edwards1968,Wohlfarth1977} These data are consistent with the result obtained from first principles calculations presented below, where the Stoner criterion has been found to be satisfied. The isothermal line in the Arrott plot passing through the origin represents the transition temperature. \cite{Ghimire2012} Thus the Curie temperature ($T_{C}$) for Cr$_{1/3}$NbS$_{2}$ is taken to be 120 K.

\subsection{Transport properties}
\begin{figure}[h]
\begin{center}
\includegraphics[scale=1.0]{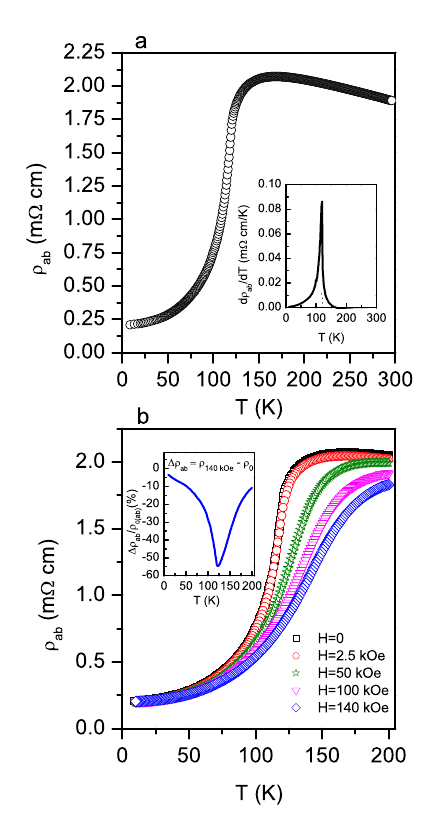}
\caption{\label{3} (Color online) a) Electrical resistivity of Cr$_{1/3}$NbS$_{2}$ as a function of temperature measured in the \emph{ab}-plane. Inset shows the temperature derivative of the resistivity. b) Temperature dependence of electrical resistivity of Cr$_{1/3}$NbS$_{2}$ measured in the \emph{ab}-plane at indicated magnetic fields applied parallel to the plane. Inset shows the magnetoresistance where $\Delta \rho_{ab}$ = $\rho_{14T}$-$\rho_{0}$.}
\end{center}
\end{figure}
Electrical resistivity as a function of temperature is depicted in Figure \ref{3}(a). Interestingly, there is an abrupt and large change of the resistivity in the vicinity of the magnetic transition temperature. Above $T_{C}$, up to the measured temperature of 300 K, there is a slight decrease in the resistivity with temperature. It contrasts with the behavior of conventional metallic ferromagnets where there is a decrease in the resistivity on cooling over the entire temperature range and a change in slope or a kink near the magnetic transition temperature due to reduction in the spin disorder scattering. Hall effect measurements at 200 K with the current in the \emph{ab}-plane and the magnetic field along the \emph{c} axis gave a Hall coefficient of 2$\times$10$^{-2}$ cm$^{3}$ C$^{-1}$ at 200 K (for a single band, this would correspond to ~10$^{20}$ holes cm$^{-3}$).  This along with the measured resistivity values near 2 m{$\Omega$ cm above the transition, suggests the material may be described as low carrier concentration metal or heavily doped semiconductor. The decrease in $\rho$ with increasing temperature at the highest temperatures investigated indicate that this is not a simple metal in the paramagnetic state. Comparison of the transport properties with those reported in Ref. \onlinecite{Parkin1980b} shows that the crystals studied here have higher resistivity and higher Hall coefficients than those previously studied.

Figure \ref{3}(b) shows the temperature dependence of the resistivity at various magnetic fields up to 140 kOe. As expected the effect of the magnetic field is strongest near the magnetic transition temperature. The sharp change observed at lower fields near $T_{C}$ are observed to skew to higher temperatures by magnetic field, as may be expected for a ferromagnet. However, a strong suppression of the resistivity upon cooling is still seen at 140 kOe. Inset of figure \ref{3}(b) shows the magnetoresistance obtained by subtracting the resistivity measured at zero applied magnetic field from the resistivity measured at the magnetic field of 140 kOe. A very large magnetoresistance of about 55 percent is observed at the Curie temperature.

 Figure \ref{4}(a) shows the magnetoresistance measured at 2 K up to an applied magnetic field of 3 kOe. It shows a maximum magnetoresistance of about 5.5 \%. The change in magnetoresistance around 1 kOe is quite sharp and coincides with the SL phase indicated by magnetization measurements. At 2 K a small hysteresis is observed [upper inset in the Figure \ref{4}(a)] and it appears in the same region where hysteresis is seen in the magnetization measurements [Fig. \ref{1}(a)]. The hysteresis persists up to 100 K (not shown). The lower inset shows the derivative of the resistivity with respect to the applied magnetic field which clearly shows the sharp change occurring  at the applied field of 1 kOe. This behavior in the resistivity also seems consistent with the soliton model. As discussed above, as the applied magnetic field starts aligning the moments towards its direction, spin disorder scattering decreases thereby reducing the electrical resistivity. As the magnetic field causes the transition from the soliton phase to the ferromagnetic state, all the moments are aligned in the direction of the field and thus no further change in the resistivity at a particular temperature.  The low field magnetoresistance behavior is found to be qualitatively similar at all temperatures measured below 100 K. Figure \ref{4}(b) shows the normalized resistivity defined by $\rho_{H}$/$\rho_{H=0}$ as a function of magnetic field. Above 120 K the resistivity varies little with field, while below 120 K the resistivity behavior is similar to that observed at 2 K [cf. Fig. \ref{4}(a)]. Interestingly, at $T_{C}$ = 120 K $\rho(H)$ has a strong field dependence but, unlike at lower temperatures, it does not saturate near 1 kOe. The origin of this behavior is unclear, and highlights the lack of detailed understanding of the complex behavior of helimagnetic materials near their critical temperatures.
\begin{figure}[h]
\begin{center}
\includegraphics[scale=1.2]{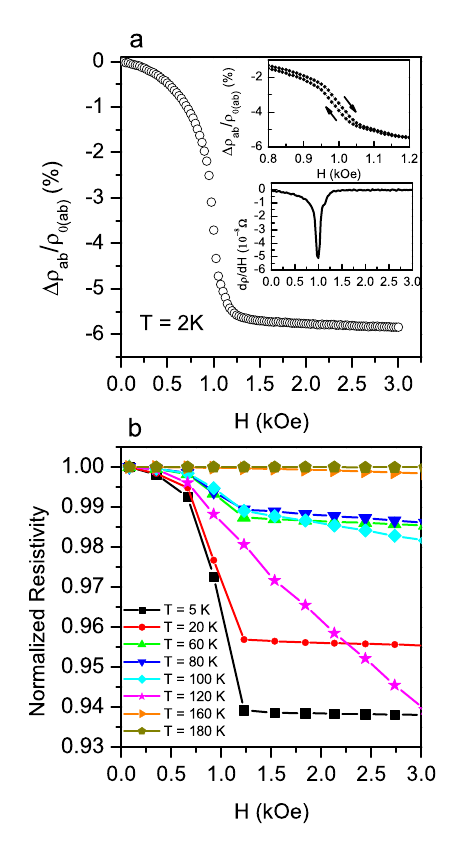}
\caption{\label{4}(Color online) Low field magnetoresistance of Cr$_{1/3}$NbS$_{2}$ measured in the \emph{ab}-plane with magnetic field applied parallel to the plane. a) Magnetoresistance measured at T = 2 K. The upper inset shows the change in slope in the magnetoresistance in the vicinity of the field where the metamagnetic transition is observed in the magnetization measurements. The lower inset shows the derivative of the resistivity with respect of the field. It shows that the sharp change occurs at 1 kOe. b) Normalized resistivity measured in the \emph{ab}-plane as a function of magnetic field applied parallel to the plane at indicated temperatures.}
\end{center}
\end{figure}

The temperature dependent thermal conductivity $\kappa$(T) of Cr$_{1/3}$NbS$_{2}$ is shown in figure \ref{7}(a). Upon heating, $\kappa$ increases up to about 65 K above which it decreases slightly up to 200 K. In general, the thermal conductivity of a metal is the sum of electronic and lattice terms. The electronic contribution to the thermal conductivity can be estimated by the Wiedemann-Franz Law \cite{kittel2005}: $\kappa_{e}\rho/T$ = \emph{L}, where $\rho$ is the electrical resistivity and \emph{L} = 2.45$\times$10$^{-8}$ W$\Omega$K$^{-2}$ is the Lorenz number. The inset in figure \ref{7}(a) shows the calculated electronic part of the thermal conductivity ($\kappa_{e}$) for Cr$_{1/3}$NbS$_{2}$. This estimation suggests that the thermal conductivity is mainly due to phonons.
\begin{figure}[h]
\begin{center}
\includegraphics[scale=1.0]{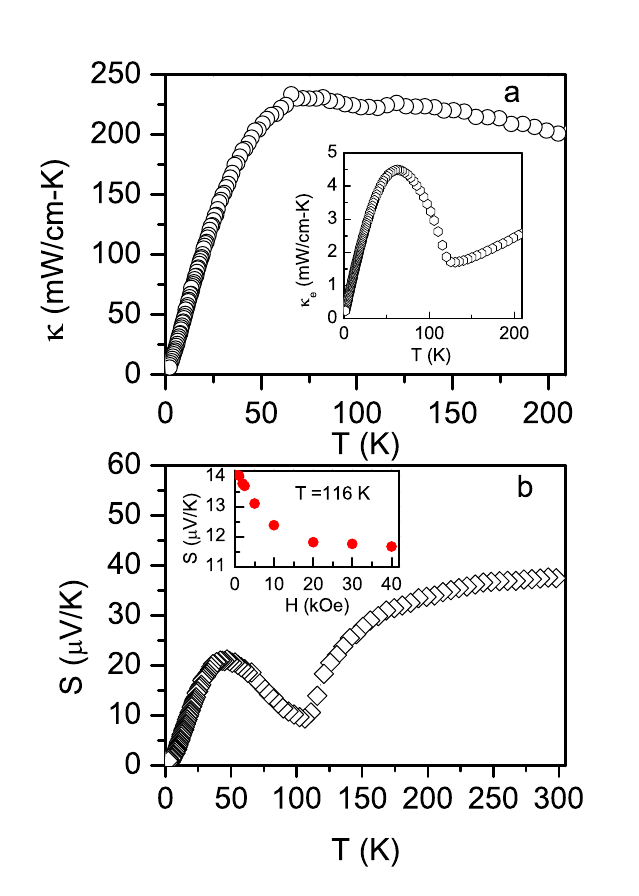}
\caption{\label{7}(Color online) a) Temperature dependence of thermal conductivity measured in the \emph{ab}-plane. Inset shows the electronic contribution to the thermal conductivity estimated from the Wiedemann-Franz law, and b) Temperature dependence of Seebeck coefficient measured in the \emph{ab}-plane. Inset shows the magnetic field dependence of the Seebeck coefficient at T = 116 K with magnetic field applied parallel to the \emph{ab}-plane}
\end{center}
\end{figure}

Figure \ref{7}(b) shows the temperature dependence of the Seebeck coefficient (\emph{S}). At all temperatures measured, the Seebeck coefficient is positive. However, it shows a remarkable minimum slightly below the magnetic transition temperature. The Seebeck coefficient is very sensitive to the details of the band structure near the Fermi energy $E_{F}$. The anomalous behavior observed near $T_{C}$ may be related to the significant decrease in the density of states (DOS) around $E_{F}$ in the magnetic state as compared to that in the non-magnetic state as presented below. Interestingly the Seebeck coefficient is found to have magnetic field dependence at 116 K. With the external magnetic field applied perpendicular to the \emph{c} axis, \emph{S} decreases up to about 20 kOe, above which it remains constant [Fig. \ref{7}(b)] which suggests either a spin-entropy contribution to \emph{S} or the effects of the magnetic field on the scattering of carriers.

\subsection{Heat capacity}
The temperature dependence of specific heat shows clear lambda anomaly in the vicinity of the magnetic ordering temperature at ambient field as shown in figure \ref{5}(a). The magnetic field dependence of the heat capacity anomaly is depicted in [Fig. \ref{5}(b)]. The lambda peak does not change significantly up to 800 Oe. However, it is found to decrease clearly at a field of 1500 Oe. The peak is suppressed with the further increase in the magnetic field and is not observed above 50 kOe. Application of a magnetic field increased the heat capacity above the magnetic transition temperature (120 K) and suppressed the heat capacity below it. A decrease in the lambda peak is found to occur in the ferromagnetic state when the moments are aligned in the direction of the magnetic field. This behavior is consistent with the results obtained in the magnetization and magnetoresistance measurements.
\begin{figure}
\begin{center}
\includegraphics[scale=1.0]{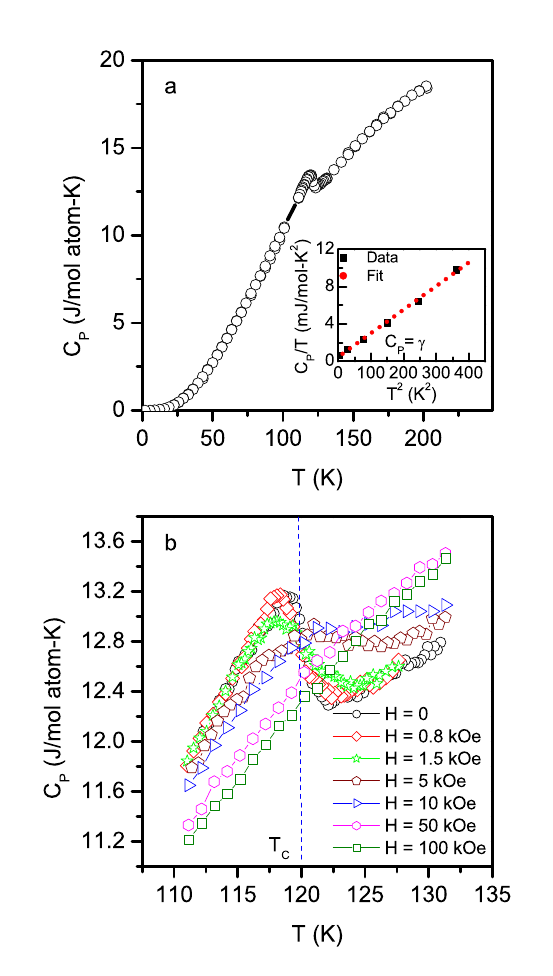}
\caption{\label{5}(Color online) a) Heat capacity of Cr$_{1/3}$NbS$_{2}$ as a function of temperature at zero applied magnetic field. b) Temperature dependence of the specific heat near the magnetic transition temperature measured at indicated magnetic fields applied perpendicular to the \emph{c} axis.}
\end{center}
\end{figure}

 The low temperature specific heat can be modeled well with electronic and phononic contributions. The inset of figure \ref{5}(a) shows a fit to \emph{C/T} = $\gamma$ + $\beta T^{2}$, where, $\gamma$ is the Sommerfeld coefficient and $\beta$ is the phononic heat capacity coefficient. The fit yielded $\gamma$ = 0.40 $\pm$ 0.25 \emph{mJ mol$^{-1}$ K$^{-2}$} and $\beta$ = 0.0026 $\pm$ 0.00047 \emph{mJ mol$^{-1}$ K$^{-4}$}. The electron density of states at the Fermi energy ($E_{F}$) and the Debye temperature ($T_{D}$) can be estimated from $\gamma$ and $\beta$ using the relations $N(E_{F})$ = 3$\gamma$/$\pi^{2}$k$_{B}^{2}$ and  $T_{D}$ = (5$\beta$/12$R$$\pi^{4}$)$^{-1/3}$, respectively, where k$_{B}$ is the Boltzmann constant and \emph{R} is the universal gas  constant. The calculated values are $N(E_{F})$ = 3.40 states (eV-u.c.)$^{-1}$ and $T_{D}$ = 419 K. The experimentally determined DOS at the Fermi energy is comparable to the value [2.76 states (eV-u.c.)$^{-1}$] obtained from the electronic structure calculations in the magnetic state presented below (Fig. \ref{10}).

\subsection{First principles calculations}

In order to understand theoretically the effect of the magnetic ordering we have performed first principles density functional theory calculations, using the generalized gradient approximation of Perdew, Burke and Ernzerhof \cite{perdew1996} as implemented in the all electron code WIEN2k. \cite{wien}  We have used the experimental hexagonal lattice constants and internal parameters; no relaxation was performed.  Calculations in both a non-magnetic state and a magnetic spin-polarized state were performed, using 120 $k$-points in the irreducible Brillouin zone and LAPW sphere radii of 2.07 a$_{0}$, 2.33 a$_{0}$ and 2.34 a$_{0}$ for the S, Nb and Cr atoms respectively (here a$_{0}$ is the Bohr radius = 0.529177 $\AA$).  An RK$_{max}$ of 7, where R is the minimum atomic sphere radius and $K$ the largest plane wave vector used in the expansion, was used.

Regarding the spin-polarized calculations, a brief discussion is in order.  As are previous experimental evidences \cite{Miyadai1983,Togawa2012} that the ground state of Cr$_{1/3}$NbS$_{2}$ is a very long wavelength ($\sim480 \AA$) spiral with adjacent spins in the spiral very nearly parallel,  such a magnetic structure, if one attempted to study it in a ``brute-force" computational framework, would result in a calculation involving several hundred atoms and several thousand electrons, which would present a nearly intractable problem for standard first principles approaches.  However, a simple observation allows one to perform a much simpler calculation which should be very close in energy and ground state properties to the actual ground state.  The origin of the Dzyaloshinsky-Moriya interaction lies in the spin-orbit interaction. \cite{Dzyaloshinsky1958, Moriya1960} The energy scale of the relativistic term believed to be responsible for the spiral state is much weaker (a quantitative estimate suggests between one and two orders of magnitude smaller) than the ordinary ferromagnetic Heisenberg nearest neighbor exchange energy.  Thus it is highly likely that the actual energies and properties of the {\it real} ground state can be described as very nearly that of an ordinary spin-polarized ferromagnetic ground state, {\it with the understanding} that the ``spin-up" and ``spin-down" description of the properties presented below, such as band structure and density-of-states, are in essence only labels and that the actual physical state is a non-collinear spiral.  It is then an excellent approximation to take the true ground state physical observable, such as the density-of-states, as the {\it sum} of the separate ``spin-up" and ``spin-down" contributions.  Again, a quantitative estimate suggests that the effect of the DM interaction on the Kohn-Sham eigenvalues is less than 10 meV, which is much smaller than all other relevant energy scales and virtually invisible on the plots presented below.

We begin with the calculated non-magnetic band structure and density-of-states, presented in Figure \ref{8} below.  There are several bands crossing the Fermi level, and in addition flat bands virtually abutting the Fermi level from M to K and H to M.  There is also a fairly complex structure around the $\Gamma$ point.  As might be expected this band structure leads to a high density-of-states at the Fermi level, presented in Figure \ref{8}(b).  The Fermi level sits near the middle of a region of fairly elevated DOS, with Fermi level DOS of approximately 15 states/eV-unit cell and approximately 10 states/eV-unit cell Chromium.  Given the two Cr in the unit cell and the exchange correlation value $I$ of 0.38 eV for Cr \cite{janak1977}, one finds the Stoner criterion $N_{0}I > 1$ well satisfied, with $N_{0}I$ on a per Cr basis having the value 1.9.  The majority, but not entirety, of the DOS character near $E_{F}$ is Cr, suggesting some hybridization of the chemical bonds.
\begin{figure}[h]
\begin{center}
\includegraphics[scale=0.8]{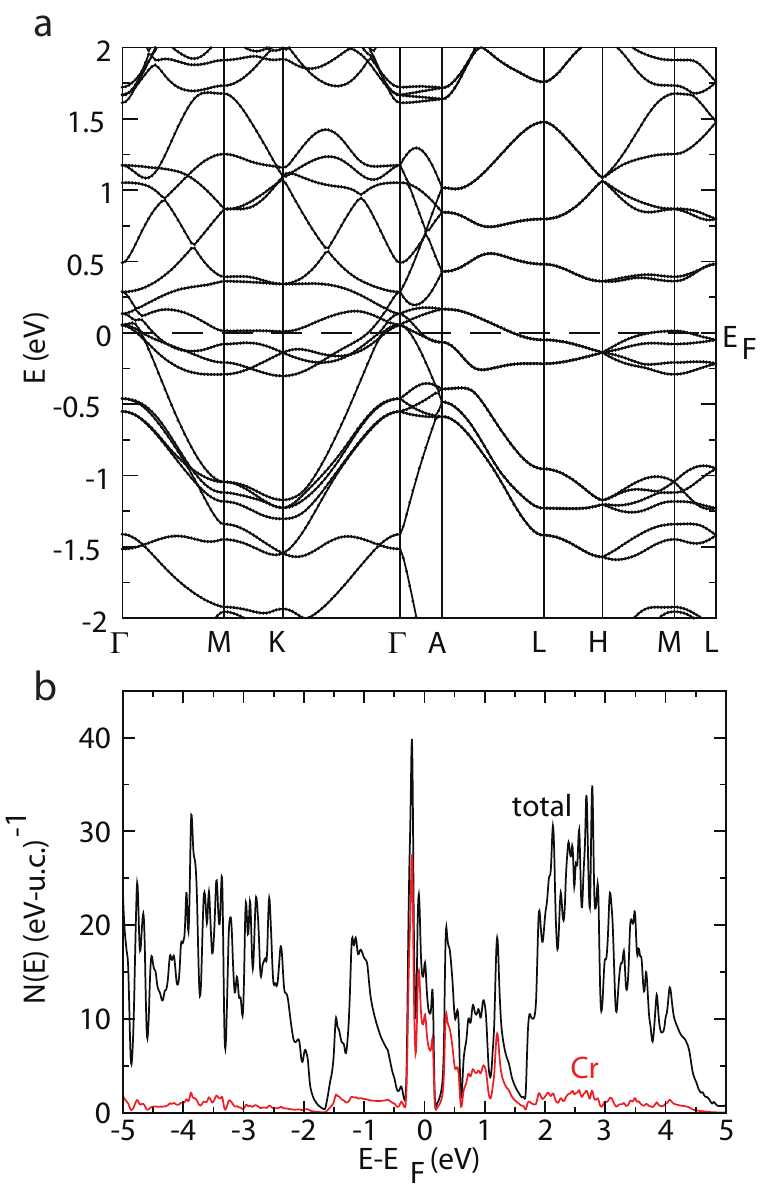}
\caption{\label{8} (Color online) a) The calculated band structure b) and density of states of Cr$_{1/3}$NbS$_{2}$ in the nonmagnetic state.  Note the many Fermi level crossings and corresponding high density of states at $E_{F}$.}
\end{center}
\end{figure}

Moving to the magnetic state calculations previously described, we find strong evidence for a magnetic ground state, with the magnetic state some 930 meV per formula unit in energy lower than the non-magnetic state.  In Figure \ref{9} we present the calculated band structure and density-of-states for this ground state; as outlined above we have simply summed the ``spin-up" and ``spin-down" contributions to the
\begin{figure}[h]
\begin{center}
\includegraphics[scale=0.8]{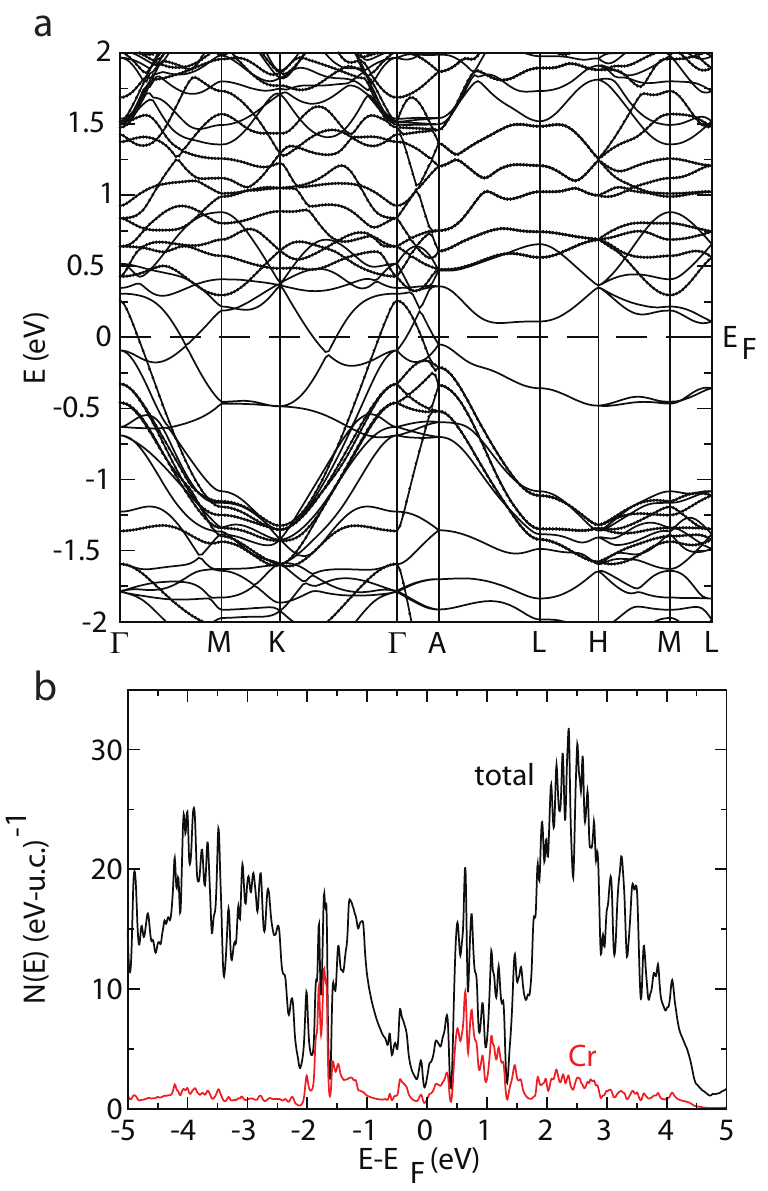}
\caption{\label{9} (Color online) a) The calculated band structure and b) density of states of Cr$_{1/3}$NbS$_{2}$ in the assumed magnetic state.  Most of the Fermi level crossings have disappeared and the density-of-states shows a substantial pseudogap at $E_{F}$.}
\end{center}
\end{figure}
DOS and plotted both band structures on the same plot.  The band structure demonstrates far fewer Fermi level crossings, and accordingly examining the density of states we find a substantial loss of spectral weight around $E_{F}$.  It is instructive to directly compare the magnetic and non-magnetic DOS and this is presented in Figure \ref{10}.  We note that, the significant changes to the DOS are confined to an energy range of $\pm$ 2 eV around $E_{F}$, despite the large energy gain of nearly an eV per formula unit. The DOS at $E_{F}$ in the magnetic state is found to be 2.76 states (eV-u.c.)$^{-1}$ which is consistent with the experimental value extracted from the low temperature heat capacity presented above.
\begin{figure}[h]
\begin{center}
\includegraphics[scale=0.8]{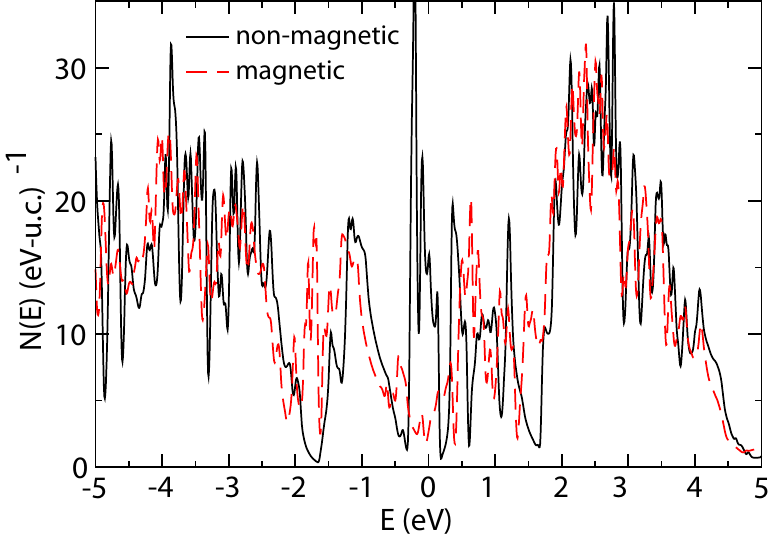}
\caption{\label{10} (Color online) The calculated density of states  of Cr$_{1/3}$NbS$_{2}$ in the magnetic and non-magnetic states.}
\end{center}
\end{figure}
\section{Summary and Conclusions}
We have determined the temperature and magnetic field dependence of the magnetic, transport, and thermal properties of Cr$_{1/3}$NbS$_{2}$ single crystals. Contours of constant magnetization are utilized to identify key features in the magnetic phase diagram near the transition region (Fig. \ref{2c}). The material is magnetically ordered below about 120 K. Below this temperature, increasing the magnetic field applied in the \emph{ab}-plane results in a transition from helimagnetic to ferromagnetic order, with evidence from the previously reported soliton lattice phase at intermediate fields. \cite{ Togawa2012} Evidence for these field- and temperature- induced phase transitions is also seen in resistivity, Seebeck coefficient, and heat capacity results. The resistivity responds strongly to the magnetic ordering, decreasing upon cooling through the magnetic transitions by more than a factor of four.  A sharp drop in the Seebeck coefficient is observed upon cooling near 120 K as well. Significant effects of applied magnetic field on these transport properties are also seen in the ordered state, especially near the ordering temperature. Observations include large magnetoresistiance (-55\% at 140 kOe and 120 K) and $\rho(H)$ data that mimic the \emph{M(H)} behavior below 120 K, but are anomalous at 120 K. These findings suggest that the magnetic transitions are accompanied by large changes in the electronic structure, which is confirmed by DFT calculations. The calculated DOS in the magnetic state is strongly suppressed relative to the value from non-magnetic calculations. Since the resistivity decreases upon entering the magnetically ordered state, changes in magnetic scattering of charge carriers are also likely to be important in determining the transport behavior. The large change in the resistivity suggests strong scattering by spin fluctuations above 120 K. Analysis of the heat capacity results indicate a small Sommerfeld coefficient of 0.4 $mJ mol^{-1} K^{-2}$, in agreement with the magnetic DFT calculations, and a Debye temperature of 419 K. A lambda-like anomaly in the heat capacity near the magnetic transition is suppressed strongly at fields above about 1.5 kOe, when the material is in the ferromagnetic state.
\section{Acknowledgements}
Research was supported by the US Department of Energy, Office of Basic Energy Sciences, Materials Sciences and Engineering Division. D. P. acknowledges support from the ORNL LDRD SEED funding project S12-006, ``Rare Earth Free Magnets: Compute, Create, Characterize".

\bibliographystyle{apsrev}


\begin{thebibliography}{30}
\expandafter\ifx\csname natexlab\endcsname\relax\def\natexlab#1{#1}\fi
\expandafter\ifx\csname bibnamefont\endcsname\relax
  \def\bibnamefont#1{#1}\fi
\expandafter\ifx\csname bibfnamefont\endcsname\relax
  \def\bibfnamefont#1{#1}\fi
\expandafter\ifx\csname citenamefont\endcsname\relax
  \def\citenamefont#1{#1}\fi
\expandafter\ifx\csname url\endcsname\relax
  \def\url#1{\texttt{#1}}\fi
\expandafter\ifx\csname urlprefix\endcsname\relax\def\urlprefix{URL }\fi
\providecommand{\bibinfo}[2]{#2}
\providecommand{\eprint}[2][]{\url{#2}}

\bibitem[{\citenamefont{M\"{u}hlbauer et~al.}(2009)\citenamefont{M\"{u}hlbauer,
  Binz, Jonietz, Pfleiderer, Rosch, Neubauer, Georgii, and
  B\"{o}ni}}]{Muhlbauer2009a}
\bibinfo{author}{\bibfnamefont{S.}~\bibnamefont{M\"{u}hlbauer}},
  \bibinfo{author}{\bibfnamefont{B.}~\bibnamefont{Binz}},
  \bibinfo{author}{\bibfnamefont{F.}~\bibnamefont{Jonietz}},
  \bibinfo{author}{\bibfnamefont{C.}~\bibnamefont{Pfleiderer}},
  \bibinfo{author}{\bibfnamefont{A.}~\bibnamefont{Rosch}},
  \bibinfo{author}{\bibfnamefont{A.}~\bibnamefont{Neubauer}},
  \bibinfo{author}{\bibfnamefont{R.}~\bibnamefont{Georgii}}, \bibnamefont{and}
  \bibinfo{author}{\bibfnamefont{P.}~\bibnamefont{B\"{o}ni}},
  \bibinfo{journal}{Science} \textbf{\bibinfo{volume}{323}},
  \bibinfo{pages}{915} (\bibinfo{year}{2009}).

\bibitem[{\citenamefont{Yu et~al.}(2011)\citenamefont{Yu, Kanazawa, Onose,
  Kimoto, Zhang, Ishiwata, Matsui, and Tokura}}]{Yu2011a}
\bibinfo{author}{\bibfnamefont{X.~Z.} \bibnamefont{Yu}},
  \bibinfo{author}{\bibfnamefont{N.}~\bibnamefont{Kanazawa}},
  \bibinfo{author}{\bibfnamefont{Y.}~\bibnamefont{Onose}},
  \bibinfo{author}{\bibfnamefont{K.}~\bibnamefont{Kimoto}},
  \bibinfo{author}{\bibfnamefont{W.~Z.} \bibnamefont{Zhang}},
  \bibinfo{author}{\bibfnamefont{S.}~\bibnamefont{Ishiwata}},
  \bibinfo{author}{\bibfnamefont{Y.}~\bibnamefont{Matsui}}, \bibnamefont{and}
  \bibinfo{author}{\bibfnamefont{Y.}~\bibnamefont{Tokura}},
  \bibinfo{journal}{Nat. Mater.} \textbf{\bibinfo{volume}{10}},
  \bibinfo{pages}{106} (\bibinfo{year}{2011}).

\bibitem[{\citenamefont{Yu et~al.}(2010)\citenamefont{Yu, Onose, Kanazawa,
  Park, Han, Matsui, Nagaosa, and Tokura}}]{Yu2010}
\bibinfo{author}{\bibfnamefont{X.~Z.} \bibnamefont{Yu}},
  \bibinfo{author}{\bibfnamefont{Y.}~\bibnamefont{Onose}},
  \bibinfo{author}{\bibfnamefont{N.}~\bibnamefont{Kanazawa}},
  \bibinfo{author}{\bibfnamefont{J.~H.} \bibnamefont{Park}},
  \bibinfo{author}{\bibfnamefont{J.~H.} \bibnamefont{Han}},
  \bibinfo{author}{\bibfnamefont{Y.}~\bibnamefont{Matsui}},
  \bibinfo{author}{\bibfnamefont{N.}~\bibnamefont{Nagaosa}}, \bibnamefont{and}
  \bibinfo{author}{\bibfnamefont{Y.}~\bibnamefont{Tokura}},
  \bibinfo{journal}{Nature} \textbf{\bibinfo{volume}{465}},
  \bibinfo{pages}{901} (\bibinfo{year}{2010}).

\bibitem[{\citenamefont{M\"{u}nzer et~al.}(2010)\citenamefont{M\"{u}nzer,
  Neubauer, Adams, M\"{u}hlbauer, Franz, Jonietz, Georgii, B\"{o}ni, Pedersen,
  Schmidt et~al.}}]{Munzer2010}
\bibinfo{author}{\bibfnamefont{W.}~\bibnamefont{M\"{u}nzer}},
  \bibinfo{author}{\bibfnamefont{A.}~\bibnamefont{Neubauer}},
  \bibinfo{author}{\bibfnamefont{T.}~\bibnamefont{Adams}},
  \bibinfo{author}{\bibfnamefont{S.}~\bibnamefont{M\"{u}hlbauer}},
  \bibinfo{author}{\bibfnamefont{C.}~\bibnamefont{Franz}},
  \bibinfo{author}{\bibfnamefont{F.}~\bibnamefont{Jonietz}},
  \bibinfo{author}{\bibfnamefont{R.}~\bibnamefont{Georgii}},
  \bibinfo{author}{\bibfnamefont{P.}~\bibnamefont{B\"{o}ni}},
  \bibinfo{author}{\bibfnamefont{B.}~\bibnamefont{Pedersen}},
  \bibinfo{author}{\bibfnamefont{M.}~\bibnamefont{Schmidt}},
  \bibnamefont{et~al.}, \bibinfo{journal}{Phys. Rev. B}
  \textbf{\bibinfo{volume}{81}}, \bibinfo{pages}{041203}
  (\bibinfo{year}{2010}).

\bibitem[{\citenamefont{Seki et~al.}(2012)\citenamefont{Seki, Yu, Ishiwata, and
  Tokura}}]{Seki2012}
\bibinfo{author}{\bibfnamefont{S.}~\bibnamefont{Seki}},
  \bibinfo{author}{\bibfnamefont{X.~Z.} \bibnamefont{Yu}},
  \bibinfo{author}{\bibfnamefont{S.}~\bibnamefont{Ishiwata}}, \bibnamefont{and}
  \bibinfo{author}{\bibfnamefont{Y.}~\bibnamefont{Tokura}},
  \bibinfo{journal}{Science} \textbf{\bibinfo{volume}{336}},
  \bibinfo{pages}{198} (\bibinfo{year}{2012}).

  \bibitem[{\citenamefont{Bogadnov et~al.}(1989)\citenamefont{Bogadnov, Kudinov,
  and Yablonskii}}]{Bogadonov1989c}
\bibinfo{author}{\bibfnamefont{A.~N.} \bibnamefont{Bogadnov}},
  \bibinfo{author}{\bibfnamefont{M.~V.} \bibnamefont{Kudinov}},
  \bibnamefont{and} \bibinfo{author}{\bibfnamefont{D.~A.}
  \bibnamefont{Yablonskii}}, \bibinfo{journal}{Sov. Phys. Solid State}
  \textbf{\bibinfo{volume}{31}}, \bibinfo{pages}{1707} (\bibinfo{year}{1989}).

\bibitem[{\citenamefont{Bogdanov et~al.}(2002)\citenamefont{Bogdanov,
  R\"{o}\ss~ler, Wolf, and M\"{u}ller}}]{Bogdanov2002}
\bibinfo{author}{\bibfnamefont{A.~N.} \bibnamefont{Bogdanov}},
  \bibinfo{author}{\bibfnamefont{U.~K.} \bibnamefont{R\"{o}\ss~ler}},
  \bibinfo{author}{\bibfnamefont{M.}~\bibnamefont{Wolf}}, \bibnamefont{and}
  \bibinfo{author}{\bibfnamefont{K.-H.} \bibnamefont{M\"{u}ller}},
  \bibinfo{journal}{Phys. Rev. B} \textbf{\bibinfo{volume}{66}},
  \bibinfo{pages}{214410} (\bibinfo{year}{2002}).

\bibitem[{\citenamefont{R\"{o}bler et~al.}(2006)\citenamefont{R\"{o}bler,
  Bogdanov, and Pfleiderer}}]{Rossler2006}
\bibinfo{author}{\bibfnamefont{U.~K.} \bibnamefont{R\"{o}$\ss$ler}},
  \bibinfo{author}{\bibfnamefont{A.~N.} \bibnamefont{Bogdanov}},
  \bibnamefont{and}
  \bibinfo{author}{\bibfnamefont{C.}~\bibnamefont{Pfleiderer}},
  \bibinfo{journal}{Nature} \textbf{\bibinfo{volume}{442}},
  \bibinfo{pages}{797} (\bibinfo{year}{2006}).

\bibitem[{\citenamefont{R\"{o}$\ss$ler et~al.}(2011)\citenamefont{R\"{o}\ss~ler,
  Leonov, and Bogdanov}}]{Rossler2011}
\bibinfo{author}{\bibfnamefont{U.~K.} \bibnamefont{R\"{o}$\ss$ler}},
  \bibinfo{author}{\bibfnamefont{A.~A.} \bibnamefont{Leonov}},
  \bibnamefont{and} \bibinfo{author}{\bibfnamefont{A.~N.}
  \bibnamefont{Bogdanov}}, \bibinfo{journal}{J. Phys: Conf. Ser.}
  \textbf{\bibinfo{volume}{303}}, \bibinfo{pages}{012105}
  (\bibinfo{year}{2011}).


\bibitem[{\citenamefont{Bak and Jensen}(1980)}]{Bak1980}
\bibinfo{author}{\bibfnamefont{P.}~\bibnamefont{Bak}} \bibnamefont{and}
  \bibinfo{author}{\bibfnamefont{M.~H.} \bibnamefont{Jensen}},
  \bibinfo{journal}{J. Phys. C: Solid St. Phys.} \textbf{\bibinfo{volume}{13}},
  \bibinfo{pages}{L881} (\bibinfo{year}{1980}).

\bibitem[{\citenamefont{Nakanishi et~al.}(1980)\citenamefont{Nakanishi, Yanase,
  Hasegawa, and Kataoka}}]{Nakanishi1980b}
\bibinfo{author}{\bibfnamefont{O.}~\bibnamefont{Nakanishi}},
  \bibinfo{author}{\bibfnamefont{A.}~\bibnamefont{Yanase}},
  \bibinfo{author}{\bibfnamefont{A.}~\bibnamefont{Hasegawa}}, \bibnamefont{and}
  \bibinfo{author}{\bibfnamefont{M.}~\bibnamefont{Kataoka}},
  \bibinfo{journal}{Solid State Commun.} \textbf{\bibinfo{volume}{35}},
  \bibinfo{pages}{995} (\bibinfo{year}{1980}).

\bibitem[{\citenamefont{Kataoka and Nakanishi}(1981)}]{Kataoka1981}
\bibinfo{author}{\bibfnamefont{M.}~\bibnamefont{Kataoka}} \bibnamefont{and}
  \bibinfo{author}{\bibfnamefont{O.}~\bibnamefont{Nakanishi}},
  \bibinfo{journal}{J. Phys. Soc. Japan} \textbf{\bibinfo{volume}{50}},
  \bibinfo{pages}{3888} (\bibinfo{year}{1981}).

\bibitem[{\citenamefont{Kataoka et~al.}(1984)\citenamefont{Kataoka, Nakanishi,
  Yanase, and Kanamori}}]{Kataoka1984}
\bibinfo{author}{\bibfnamefont{M.}~\bibnamefont{Kataoka}},
  \bibinfo{author}{\bibfnamefont{O.}~\bibnamefont{Nakanishi}},
  \bibinfo{author}{\bibfnamefont{A.}~\bibnamefont{Yanase}}, \bibnamefont{and}
  \bibinfo{author}{\bibfnamefont{J.}~\bibnamefont{Kanamori}},
  \bibinfo{journal}{J. Phys. Soc. Japan} \textbf{\bibinfo{volume}{53}},
  \bibinfo{pages}{3624} (\bibinfo{year}{1984}).

\bibitem[{\citenamefont{Bogdanov and Yablonskii}(1989)}]{Bogdanov1989}
\bibinfo{author}{\bibfnamefont{A.~N.} \bibnamefont{Bogdanov}} \bibnamefont{and}
  \bibinfo{author}{\bibfnamefont{D.~A.} \bibnamefont{Yablonskii}},
  \bibinfo{journal}{Sov. Phys. JEPT} \textbf{\bibinfo{volume}{68}},
  \bibinfo{pages}{101} (\bibinfo{year}{1989}).

\bibitem[{\citenamefont{Bogdanov and Hubert}(1994)}]{Bogdanov1994}
\bibinfo{author}{\bibfnamefont{A.}~\bibnamefont{Bogdanov}} \bibnamefont{and}
  \bibinfo{author}{\bibfnamefont{A.}~\bibnamefont{Hubert}},
  \bibinfo{journal}{J. Magn. Magn. Mater} \textbf{\bibinfo{volume}{138}},
  \bibinfo{pages}{255} (\bibinfo{year}{1994}).

  \bibitem[{\citenamefont{Hamann et~al.}(2011)\citenamefont{Hamann, Lamago, Wolf,
  L\"{o}hneysen, and Reznik}}]{Hamann2011}
\bibinfo{author}{\bibfnamefont{A.}~\bibnamefont{Hamann}},
  \bibinfo{author}{\bibfnamefont{D.}~\bibnamefont{Lamago}},
  \bibinfo{author}{\bibfnamefont{T.}~\bibnamefont{Wolf}},
  \bibinfo{author}{\bibfnamefont{H.~v.} \bibnamefont{L\"{o}hneysen}},
  \bibnamefont{and} \bibinfo{author}{\bibfnamefont{D.}~\bibnamefont{Reznik}},
  \bibinfo{journal}{Phys. Rev. Lett.} \textbf{\bibinfo{volume}{107}},
  \bibinfo{pages}{037207} (\bibinfo{year}{2011}).

\bibitem[{\citenamefont{Fischer et~al.}(2008)\citenamefont{Fischer, Shah, and
  Rosch}}]{Fischer2008}
\bibinfo{author}{\bibfnamefont{I.}~\bibnamefont{Fischer}},
  \bibinfo{author}{\bibfnamefont{N.}~\bibnamefont{Shah}}, \bibnamefont{and}
  \bibinfo{author}{\bibfnamefont{A.}~\bibnamefont{Rosch}},
  \bibinfo{journal}{Phys. Rev. B} \textbf{\bibinfo{volume}{77}},
  \bibinfo{pages}{024415} (\bibinfo{year}{2008}).

\bibitem[{\citenamefont{Wright and Mermin}(1989)}]{Wright1989}
\bibinfo{author}{\bibfnamefont{D.~C.}~\bibnamefont{Wright}} \bibnamefont{and}
  \bibinfo{author}{\bibfnamefont{N.~D.}~\bibnamefont{Mermin}},
  \bibinfo{journal}{Rev. Mod. Phys.} \textbf{\bibinfo{volume}{61}},
  \bibinfo{pages}{385} (\bibinfo{year}{1989}).

\bibitem[{\citenamefont{{Van Laar}}(1971)}]{VanLaar1971a}
\bibinfo{author}{\bibfnamefont{B.}~\bibnamefont{{Van Laar}}},
  \bibinfo{journal}{J. Solid State Chem.} \textbf{\bibinfo{volume}{3}},
  \bibinfo{pages}{154} (\bibinfo{year}{1971}).

\bibitem[{\citenamefont{Hulligert and Pobitschka}(1970)}]{Hulliger1970}
\bibinfo{author}{\bibfnamefont{F.}~\bibnamefont{Hulliger}} \bibnamefont{and}
  \bibinfo{author}{\bibfnamefont{E.~V.~A.} \bibnamefont{Pobitschka}},
  \bibinfo{journal}{J. Solid State Chem.} \textbf{\bibinfo{volume}{1}},
  \bibinfo{pages}{117} (\bibinfo{year}{1970}).

\bibitem[{\citenamefont{Miyadai et~al.}(1983)\citenamefont{Miyadai, Kikuchi,
  Kondo, Sakka, Arai, and Ishikawa}}]{Miyadai1983}
\bibinfo{author}{\bibfnamefont{T.}~\bibnamefont{Miyadai}},
  \bibinfo{author}{\bibfnamefont{K.}~\bibnamefont{Kikuchi}},
  \bibinfo{author}{\bibfnamefont{H.}~\bibnamefont{Kondo}},
  \bibinfo{author}{\bibfnamefont{S.}~\bibnamefont{Sakka}},
  \bibinfo{author}{\bibfnamefont{K.}~\bibnamefont{Arai}}, \bibnamefont{and}
  \bibinfo{author}{\bibfnamefont{Y.}~\bibnamefont{Ishikawa}},
  \bibinfo{journal}{J. Phys. Soc. Japan} \textbf{\bibinfo{volume}{52}},
  \bibinfo{pages}{1394} (\bibinfo{year}{1983}).

\bibitem[{\citenamefont{Parkin and Friend}(1980{\natexlab{a}})}]{Parkin1980b}
\bibinfo{author}{\bibfnamefont{S.~S.~P.} \bibnamefont{Parkin}}
  \bibnamefont{and} \bibinfo{author}{\bibfnamefont{R.~H.}
  \bibnamefont{Friend}}, \bibinfo{journal}{Philosophical Magazine B}
  \textbf{\bibinfo{volume}{41}}, \bibinfo{pages}{96}
  (\bibinfo{year}{1980}{\natexlab{a}}).

\bibitem[{\citenamefont{Parkin and Friend}(1980{\natexlab{b}})}]{Parkin1980c}
\bibinfo{author}{\bibfnamefont{S.~S.~P.} \bibnamefont{Parkin}}
  \bibnamefont{and} \bibinfo{author}{\bibfnamefont{R.~H.}
  \bibnamefont{Friend}}, \bibinfo{journal}{Physica B + C}
  \textbf{\bibinfo{volume}{99}}, \bibinfo{pages}{219}
  (\bibinfo{year}{1980}{\natexlab{b}}).

\bibitem[{\citenamefont{Togawa et~al.}(2012)\citenamefont{Togawa, Koyama,
  Takayanagi, Mori, Kousaka, Akimitsu, Nishihara, Inoue, Ovchinnikov, and
  Kishine}}]{Togawa2012}
\bibinfo{author}{\bibfnamefont{Y.}~\bibnamefont{Togawa}},
  \bibinfo{author}{\bibfnamefont{T.}~\bibnamefont{Koyama}},
  \bibinfo{author}{\bibfnamefont{K.}~\bibnamefont{Takayanagi}},
  \bibinfo{author}{\bibfnamefont{S.}~\bibnamefont{Mori}},
  \bibinfo{author}{\bibfnamefont{Y.}~\bibnamefont{Kousaka}},
  \bibinfo{author}{\bibfnamefont{J.}~\bibnamefont{Akimitsu}},
  \bibinfo{author}{\bibfnamefont{S.}~\bibnamefont{Nishihara}},
  \bibinfo{author}{\bibfnamefont{K.}~\bibnamefont{Inoue}},
  \bibinfo{author}{\bibfnamefont{A.~S.}~\bibnamefont{Ovchinnikov}},
  \bibnamefont{and} \bibinfo{author}{\bibfnamefont{J.}~\bibnamefont{Kishine}},
  \bibinfo{journal}{Phys. Rev. Lett.} \textbf{\bibinfo{volume}{108}},
  \bibinfo{pages}{107202} (\bibinfo{year}{2012}).

  \bibitem[{\citenamefont{Bostrem et~al.}(2008)\citenamefont{Bostrem, Kishine,
  and Ovchinnikov}}]{Bostrem2008}
\bibinfo{author}{\bibfnamefont{I.~G.} \bibnamefont{Bostrem}},
  \bibinfo{author}{\bibfnamefont{J.-I.} \bibnamefont{Kishine}},
  \bibnamefont{and} \bibinfo{author}{\bibfnamefont{A.~S.}
  \bibnamefont{Ovchinnikov}}, \bibinfo{journal}{Phys. Rev. B}
  \textbf{\bibinfo{volume}{78}}, \bibinfo{pages}{064425}
  (\bibinfo{year}{2008}).

\bibitem[{\citenamefont{Kishine and Ovchinnikov}(2009)}]{Kishine2009}
\bibinfo{author}{\bibfnamefont{J.-I.} \bibnamefont{Kishine}} \bibnamefont{and}
  \bibinfo{author}{\bibfnamefont{A.~S.} \bibnamefont{Ovchinnikov}},
  \bibinfo{journal}{Phys. Rev. B} \textbf{\bibinfo{volume}{79}},
  \bibinfo{pages}{220405(R)} (\bibinfo{year}{2009}).

\bibitem[{\citenamefont{Kishine et~al.}(2011)\citenamefont{Kishine, Proskurin,
  and Ovchinnikov}}]{Kishine2011}
\bibinfo{author}{\bibfnamefont{J.-I.} \bibnamefont{Kishine}},
  \bibinfo{author}{\bibfnamefont{I.~V.} \bibnamefont{Proskurin}},
  \bibnamefont{and} \bibinfo{author}{\bibfnamefont{A.~S.}
  \bibnamefont{Ovchinnikov}}, \bibinfo{journal}{Phys. Rev. Lett.}
  \textbf{\bibinfo{volume}{107}}, \bibinfo{pages}{017205}
  (\bibinfo{year}{2011}).

\bibitem[{\citenamefont{Kiselev et~al.}(2011)\citenamefont{Kiselev, Bogdanov,
  Sch\"{a}fer, and R\"{o}$\ss$ler}}]{Kiselev2011}
\bibinfo{author}{\bibfnamefont{N.~S.} \bibnamefont{Kiselev}},
  \bibinfo{author}{\bibfnamefont{A.~N.} \bibnamefont{Bogdanov}},
  \bibinfo{author}{\bibfnamefont{R.}~\bibnamefont{Sch\"{a}fer}},
  \bibnamefont{and} \bibinfo{author}{\bibfnamefont{U.~K.}
  \bibnamefont{R\"{o}$\ss$ler}}, \bibinfo{journal}{J. Phys. D: Appl. Phys.}
  \textbf{\bibinfo{volume}{44}}, \bibinfo{pages}{392001}
  (\bibinfo{year}{2011}).


\bibitem[{\citenamefont{Moriya and Miyadai}(1982)}]{Moriya1982}
\bibinfo{author}{\bibfnamefont{T.}~\bibnamefont{Moriya}} \bibnamefont{and}
  \bibinfo{author}{\bibfnamefont{T.}~\bibnamefont{Miyadai}},
  \bibinfo{journal}{Solid State Commun.} \textbf{\bibinfo{volume}{42}},
  \bibinfo{pages}{209} (\bibinfo{year}{1982}).

\bibitem[{\citenamefont{Beal}(1979)}]{Beal1979}
\bibinfo{author}{\bibfnamefont{A.~R.} \bibnamefont{Beal}}, in
  \emph{\bibinfo{booktitle}{Intercalated Layered Materials}}, edited by
  \bibinfo{editor}{\bibfnamefont{F.~A.} \bibnamefont{Levy}}
  (\bibinfo{publisher}{D. Reidel Publishing Company},
  \bibinfo{address}{Dordecht, Holland}, \bibinfo{year}{1979}), pp.
  \bibinfo{pages}{251--305}.

\bibitem[{\citenamefont{Parkin and Friend}(1980{\natexlab{c}})}]{Parkin1980a}
\bibinfo{author}{\bibfnamefont{S.~S.~P.} \bibnamefont{Parkin}}
  \bibnamefont{and} \bibinfo{author}{\bibfnamefont{R.~H.}
  \bibnamefont{Friend}}, \bibinfo{journal}{Philosophical Magazine B}
  \textbf{\bibinfo{volume}{41}}, \bibinfo{pages}{65}
  (\bibinfo{year}{1980}{\natexlab{c}}).

\bibitem[{\citenamefont{Chattopadhyay et~al.}(2009)\citenamefont{Chattopadhyay,
  Arora, and Roy}}]{Chattopadhyay2009}
\bibinfo{author}{\bibfnamefont{M.~K.} \bibnamefont{Chattopadhyay}},
  \bibinfo{author}{\bibfnamefont{P.}~\bibnamefont{Arora}}, \bibnamefont{and}
  \bibinfo{author}{\bibfnamefont{S.~B.} \bibnamefont{Roy}},
  \bibinfo{journal}{J. Phys.:Condens. Matter.} \textbf{\bibinfo{volume}{21}},
  \bibinfo{pages}{296003} (\bibinfo{year}{2009}).

\bibitem[{\citenamefont{Ludgren et~al.}(1970)\citenamefont{Ludgren, Beckman,
  Attia, Bhattacheriee, and Richardson}}]{Ludgren1970}
\bibinfo{author}{\bibfnamefont{L.}~\bibnamefont{Ludgren}},
  \bibinfo{author}{\bibfnamefont{O.}~\bibnamefont{Beckman}},
  \bibinfo{author}{\bibfnamefont{V.}~\bibnamefont{Attia}},
  \bibinfo{author}{\bibfnamefont{S.~P.} \bibnamefont{Bhattacheriee}},
  \bibnamefont{and}
  \bibinfo{author}{\bibfnamefont{M.}~\bibnamefont{Richardson}},
  \bibinfo{journal}{Phys. Scripta.} \textbf{\bibinfo{volume}{1}},
  \bibinfo{pages}{69} (\bibinfo{year}{1970}).

\bibitem[{\citenamefont{Onose et~al.}(2005)\citenamefont{Onose, Takeshita,
  Terakura, Takagi, and Tokura}}]{Onose2005}
\bibinfo{author}{\bibfnamefont{Y.}~\bibnamefont{Onose}},
  \bibinfo{author}{\bibfnamefont{N.}~\bibnamefont{Takeshita}},
  \bibinfo{author}{\bibfnamefont{C.}~\bibnamefont{Terakura}},
  \bibinfo{author}{\bibfnamefont{H.}~\bibnamefont{Takagi}}, \bibnamefont{and}
  \bibinfo{author}{\bibfnamefont{Y.}~\bibnamefont{Tokura}},
  \bibinfo{journal}{Phys. Rev. B} \textbf{\bibinfo{volume}{72}},
  \bibinfo{pages}{224431} (\bibinfo{year}{2005}).

\bibitem[{\citenamefont{Arrott and Noakes}(1967)}]{Arrott1967}
\bibinfo{author}{\bibfnamefont{A.}~\bibnamefont{Arrott}} \bibnamefont{and}
  \bibinfo{author}{\bibfnamefont{J.~E.} \bibnamefont{Noakes}},
  \bibinfo{journal}{Phys. Rev. Lett.} \textbf{\bibinfo{volume}{19}},
  \bibinfo{pages}{786} (\bibinfo{year}{1967}).

\bibitem[{\citenamefont{Edwards and Wohlfarth}(1968)}]{Edwards1968}
\bibinfo{author}{\bibfnamefont{D.~M.} \bibnamefont{Edwards}} \bibnamefont{and}
  \bibinfo{author}{\bibfnamefont{E.~P.} \bibnamefont{Wohlfarth}},
  \bibinfo{journal}{Proc. Roy. Soc. A.} \textbf{\bibinfo{volume}{303}},
  \bibinfo{pages}{127} (\bibinfo{year}{1968}).

\bibitem[{\citenamefont{Wohlfarth}(1977)}]{Wohlfarth1977}
\bibinfo{author}{\bibfnamefont{E.~P.} \bibnamefont{Wohlfarth}},
  \bibinfo{journal}{Physica B + C} \textbf{\bibinfo{volume}{91}},
  \bibinfo{pages}{305} (\bibinfo{year}{1977}).

\bibitem[{\citenamefont{Ghimire et~al.}(2012)\citenamefont{Ghimire, McGuire,
  Parker, Sales, Yan, Keppens, Koehler, Latture, and Mandrus}}]{Ghimire2012}
\bibinfo{author}{\bibfnamefont{N.~J.} \bibnamefont{Ghimire}},
  \bibinfo{author}{\bibfnamefont{M.~A.} \bibnamefont{McGuire}},
  \bibinfo{author}{\bibfnamefont{D.~S.} \bibnamefont{Parker}},
  \bibinfo{author}{\bibfnamefont{B.~C.} \bibnamefont{Sales}},
  \bibinfo{author}{\bibfnamefont{J.-Q.} \bibnamefont{Yan}},
  \bibinfo{author}{\bibfnamefont{V.}~\bibnamefont{Keppens}},
  \bibinfo{author}{\bibfnamefont{M.}~\bibnamefont{Koehler}},
  \bibinfo{author}{\bibfnamefont{R.~M.} \bibnamefont{Latture}},
  \bibnamefont{and} \bibinfo{author}{\bibfnamefont{D.}~\bibnamefont{Mandrus}},
  \bibinfo{journal}{Phys. Rev. B} \textbf{\bibinfo{volume}{85}},
  \bibinfo{pages}{224405} (\bibinfo{year}{2012}) and references therein.

\bibitem[{\citenamefont{Kittel}(2005)}]{kittel2005}
\bibinfo{author}{\bibfnamefont{C.}~\bibnamefont{Kittel}},
  \emph{\bibinfo{title}{Introduction to Solid State Physics}}
  (\bibinfo{publisher}{John Wiley \& Sons, Inc}, \bibinfo{address}{USA},
  \bibinfo{year}{2005}), \bibinfo{edition}{8th} ed.

\bibitem[{\citenamefont{Perdew et~al.}(1996)\citenamefont{Perdew, Burke, and
  Ernzerhof}}]{perdew1996}
\bibinfo{author}{\bibfnamefont{J.~P.}~\bibnamefont{Perdew}},
  \bibinfo{author}{\bibfnamefont{K.}~\bibnamefont{Burke}}, \bibnamefont{and}
  \bibinfo{author}{\bibfnamefont{M.}~\bibnamefont{Ernzerhof}},
  \bibinfo{journal}{Phys. Rev. Lett.} \textbf{\bibinfo{volume}{77}},
  \bibinfo{pages}{3865} (\bibinfo{year}{1996}).

\bibitem[{\citenamefont{Blaha et~al.}(2001)\citenamefont{Blaha, Schwarz,
  Madsen, Kvasnicka, and Luitz}}]{wien}
\bibinfo{author}{\bibfnamefont{P.}~\bibnamefont{Blaha}},
  \bibinfo{author}{\bibfnamefont{K.}~\bibnamefont{Schwarz}},
  \bibinfo{author}{\bibfnamefont{G.~K.~H.} \bibnamefont{Madsen}},
  \bibinfo{author}{\bibfnamefont{D.}~\bibnamefont{Kvasnicka}},
  \bibnamefont{and} \bibinfo{author}{\bibfnamefont{J.}~\bibnamefont{Luitz}},
  \emph{\bibinfo{title}{{WIEN2k, An Augmented Plane Wave + Local Orbitals
  Program for Calculating Crystal Properties}}} (\bibinfo{publisher}{Karlheinz
  Schwarz, Techn. Universitat Wein}, \bibinfo{address}{Austria},
  \bibinfo{year}{2001}).

\bibitem[{\citenamefont{Dzaloshinsky}(1958)}]{Dzyaloshinsky1958}
\bibinfo{author}{\bibfnamefont{I.}~\bibnamefont{Dzaloshinsky}},
  \bibinfo{journal}{J. Phys. Chem. Solids} \textbf{\bibinfo{volume}{4}},
  \bibinfo{pages}{241} (\bibinfo{year}{1958}).

\bibitem[{\citenamefont{Moriya}(1960)}]{Moriya1960}
\bibinfo{author}{\bibfnamefont{T.}~\bibnamefont{Moriya}},
  \bibinfo{journal}{Phys. Rev.} \textbf{\bibinfo{volume}{120}},
  \bibinfo{pages}{91} (\bibinfo{year}{1960}).

\bibitem[{\citenamefont{Janak}(1977)}]{janak1977}
\bibinfo{author}{\bibfnamefont{J.~F.} \bibnamefont{Janak}},
  \bibinfo{journal}{Phys. Rev. B} \textbf{\bibinfo{volume}{16}},
  \bibinfo{pages}{255} (\bibinfo{year}{1977}).

\end{thebibliography}

\end{document}